# METADATA CHALLENGE FOR QUERY PROCESSING OVER HETEROGENEOUS WIRELESS SENSOR NETWORK

C.Komalavalli[1], Chetna Laroiya[2]

[1]Assoc.Prof, Department of MCA, Jagan Institute of Management Studies, Rohini, New Delhi

`komal@jimsindia.org`

[2]Asst.Prof, Department of MCA, Jagan Insitute of Management Studies, Rohini, New Delhi

`chetnalaroiya@jimsindia.org`

## ABSTRACT

*Wireless sensor networks become integral part of our life. These networks can be used for monitoring the data in various domain due to their flexibility and functionality. Query processing and optimization in the WSN is a very challenging task because of their energy and memory constraint. In this paper, first our focus is to review the different approaches that have significant impacts on the development of query processing techniques for WSN. Finally, we aim to illustrate the existing approach in popular query processing engines with future research challenges in query optimization.*

## Keywords

*WSN, Query processing, Query optimization, Base Station, Cluster Node*

## 1. INTRODUCTION

WSN consists of set of small devices known as sensors which comprises of processor, memory, transducers and low range radio transreceiver. These are powered by small batteries .These sensors have capability to sense the surroundings and can perform some limited computation. Application of this includes military, traffic surveillance, target tracking, environment monitoring, healthcare monitoring, patient monitoring etc..Each node in the network is capable of monitoring the environment and storing the observed values. These values are processed and exchanged with other nodes over the wireless network. Users can query the environment through the base station to get aggregated data from the nodes. The queries can include the operators such as selection, projection, union and aggregation. Advanced research has been conducted in the two query processing engines TinyDB and Cougar. TinyDB is not supporting the join queries but supports storage point.

The rest of the paper is organized as follows. Section 1 includes query processing definition and related work. Section 2 describes about the query processing requirements, Section 3 describes about the various challenges posed in query processing and optimization and our solution to those problems. Finally Section 4 concludes and open research issues in this field.

## 1.1 QUERY PROCESSING DEFINITION

Many WSN applications are deployed to monitor the physical world by querying and analyzing the sensor data. Sensor network database stores the sensor data such as temperature, light, sound etc. and stores the characteristics of the sensor such as location, type of sensor etc. Users can issue declarative queries without having to worry about how the data is generated, query is processed, and response is transferred within the network.

Each and every sensor of a network generates tuples broadly known as the source. The collection of these tuples creates a snapshot. Every tuple of the temperature sensor consists of the information about the node location, timestamp and temperature and the acoustic sensor generates the tuple in the form of node location, timestamp, vehicle type and detection confidence.[13]

The snapshot is horizontally partitioned across the sensors in the group. The tuples generated by the temperature sensor constitutes the temperature table which is a virtual table in the wireless sensor world. These tables represent the streams of data. Querying these virtual tables are converted into corresponding data collecting operations such as get temperature and light intensity.[13]

The WSN should be able to concurrently handle several user requests through running multiple queries since the queries are coming from different WSN application in a distributed manner. The query can retrieve the result from heterogeneous WSN e.g. temperature sensing WSN, pressure sensing network, weather and traffic control WSN. So the query processing must be capable of handling these type of issues. Query engines and operators are able to adjust their behavior according to the constraints of the WSN.

## 1.2 RELATED WORK

Extensive study have been done in the field of query processing of wireless sensor network. Query processing of traditional databases are not suitable for WSN because of its failures, resource limitations in terms of energy or computation power or memory size, data streaming and mobility of the sensor, longevity of network. So we need an efficient strategy for handling distributed processing of simple queries with join operation.

The study of most sensor applications are based on a centralized approach with base station for managing the collected data and the base station generates the query plan according to the operational cost. The gateway receives the queries and forward that query to the network according to the dissemination schemes. Within the network, the query has to be processed and reports the result back to the base station. The previous studies discusses about the network aggregation and cross layer optimizations.

Query optimization reduces the number of queries injected into the wireless network in order to lower the overall traffic and resource usage of the network. Existing studies in this area reveals that user queries are received at the base station has to be optimized before disseminated into the network. But they have not considered the role of metadata in this optimization.

In certain existing query processing systems such as COUGAR1) and TinyDB7), network statistics (or network metadata) such as patterns of data produced by individual nodes, location information and energy reserves of nodes, etc. are sent back periodically to the central node (server) which originally injected a query into the network. Using the centrally collected data, the server, which
now has a detailed overview of the status of the entire network of nodes, calculates the optimal method in which the query may be evaluated. So the server generates a set of instructions that are then sent out to the individual nodes explaining the role individual nodes will play in evaluating the query, e.g. the server may stipulate which specific nodes would be required to perform aggregation
of data for a particular query. [20] Each node in TinyDB maintains a catalog of metadata that describes its local attributes, events, and user-defined functions. This metadata is periodically copied to the root of the network for use by the optimizer. Metadata are registered with the system via static linking done at compile time using the TinyOS C-like programming language. [21]

Centralized management of metadata can only optimize centralized query processing but to support in-network query processing metadata at sensor node is also required. Following challenges in metadata management are stated by scholars:

1) For a WSN, metadata should be distributed through the network
2) Metadata collection: Base station need to collect metadata from sensor nodes and calculate those metadata to work out new metadata describing the global states of the WSN.
3) Cost consideration.: Metadata management also consumes energy, example, collecting metadata from nodes at sink consumes the energy of nodes.

Research work on global metadata management was already done by the researchers. That model was not discussing about the attributes at different level. We state a metadata management model with discretion between attributes at global (sink node) and at local (cluster head) level. Our research work also aligned global metadata with in network query partitioning using our concepts of Reactive on Demand Routing Protocol.

## 2. REQUIREMENTS FOR QUERY PROCESSING

### 2.1 QUERY OPTIMIZATION: TRADITIONAL DBMS VS. UDBMS

For each query multiple query plans are possible in such a way that each plan shows a different path for retrieving the data. Selecting the efficient query plan is known as query optimization. In traditional databases, it is based upon the minimum number of disk assesses.

But in the case of WSN, a query plan has to be selected according to the energy cost which comprises of two components namely Sensing cost and Communication cost. Most wireless sensor applications include sensor data with stored data. Sensing cost includes the periodical sampling at the nodes and Communication cost includes the routing the queries to the nodes and collecting the results back from the network. A query plan must include the query execution including routing, sensing and collection of data and metadata.

In WSN, we are following two approaches for processing sensor queries: 1) distributed approach and 2) warehousing approach

In the first one, the query workload will be determined by the amount of data that should be extracted from the sensor network. In this approach, sensor query can be evaluated at the front end server / sink, at the sensors, in the network and combination of all three. [11]

In the mining approach, extracted data will be stored in a database server and query processing takes place in that database only. This approach is suitable for historical data.
Both TinyDB and Cougar supports tree based topology for query processing.

**2.2 QUERY GENERATION, DISSEMINATION & RESPONSE TRANSFER :**

In WSNs, user queries are received at the sink/gateway node which then coordinates the query processing tasks within the network. Such a centralized query generation and limited scope processing scheme does not fit to wireless sensor network. where the query source can be any node in any WSN. Moreover, instead of being limited within a WSN a query may be processed across a number of heterogeneous WSNs. We term a query as local query if the scope of a query covers the sensors within a single WSN; otherwise the query is a global query. [2]

Sensor queries involve stored data and sensor data, i.e. relations and sequences. We define a sensor query as an acyclic graph of relational and sequence operators. The inputs of a relational operator are base relations or the output of another relational operator; the inputs of a sequence operator are base sequences or the output of another sequence operator, i.e. relations are manipulated using relational operators and sequences are manipulated using sequence operators.

**2.2.1 Assumptions**

A group of WSN where each WSN includes in sensor nodes and base station taken as server is considered in our study with the following assumptions

- Each WSN is considered homogeneous as all of the sensor nodes have the same
initial energy, UDBMS, operating system, communication protocol..

- One WSN network is homogeneous because within WSN we don't have different classes of nodes.

- Each sensor nodes can operate either in sensing mode to monitor the environment parameters and transmit to the base station or can gather data, aggregate it and forward to the fusion node.

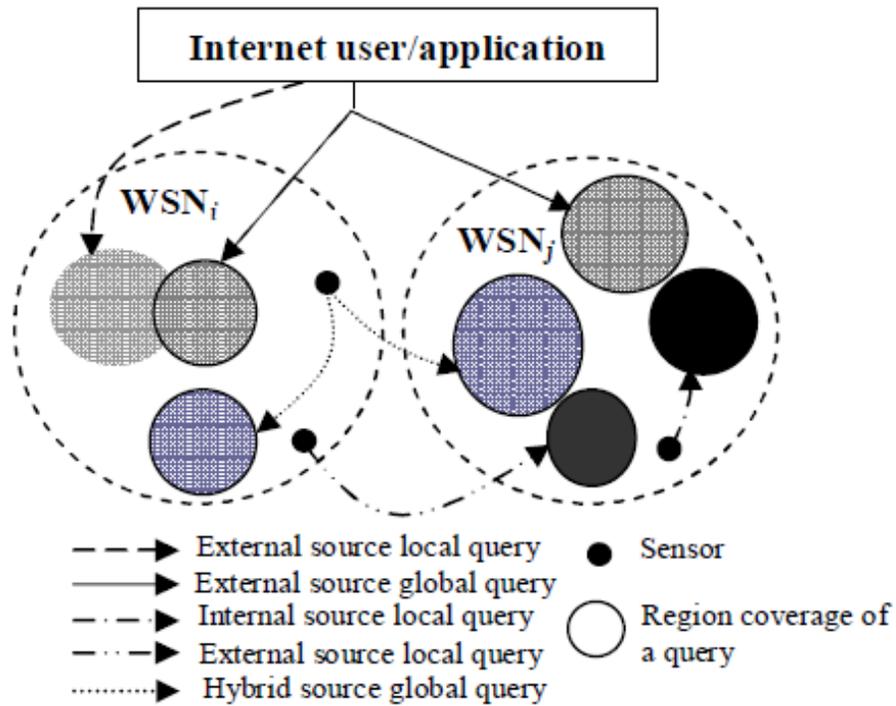

**Fig 1: Query source & scope [2]**

- A query source can only be external to WSN (sink node) not internal to WSN, with local or global query. As per Fig1 we do not consider internal source local query, external source local query and hybrid source global query.

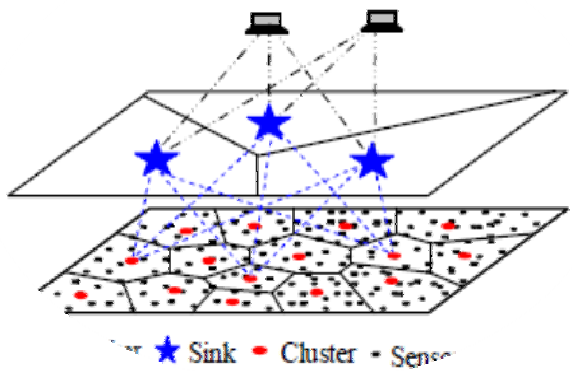

**Fig 2: A multisink clustered WSN model [2]**

We elaborate the model stated in [2] to discuss the challenges of query processing and optimization.

**Example query**

select region, count (*)
from sensors
where sound>450db and temp>45$^0$ C
group by region
epoch duration 1000s

We need to count number of sensors in each region at sound level > 450db and temperature > 45$^0$C

## 3. QUERY PROCESSING ISSUES AND RESEARCH CHALLENGES

### 3.1 ISSUE: METADATA

*Metadata* is statistics describing the distribution of the values generated by the sensors and it plays an important role in query optimization for database management systems. Metadata that describes the local attributes, events and user defined functions are managed in the each node in TinyDB. This data is updated in the sink node for the query optimization.

To generate a good plan for a user query, the optimizer requires metadata about the status of the sensor network to evaluate the costs acid benefits (latency and accuracy) of different plans. A catalog could be built and maintained at the server to maintain important information, like sensor position (potentially aggregated), density and connectivity, system workload, and network stability. System generate queries could be used to update the catalog periodically, or the catalog could be assembled dynamically through gossip-style information dissemination.

Due to the size of the metadata and the dynamics of the sensor network, it is likely prohibitive to collect all metadata at a central node, and to keep them always sufficiently up-to-date. However, energy overhead to collect metadata from sensor node to the base station is significant. However, if metadata is not collected, the query optimizer may not accurately estimate the energy cost of query plans and query plan is not ensured to be energy efficient.

**RESEARCH CHALLENGE:**

To select a good plan for a given query, the optimizer requires details about the status of the sensor network to evaluate the costs of different plans. A catalog could be built and maintained at the base station or sink to maintain network information, like sensor position, density, connectivity. System generate queries could be used to update the catalog periodically, or the catalog could be updated dynamically through brute-force information dissemination. Due to the size of the metadata and the dynamics of the sensor network, it is likely not desired to maintain all metadata at a central node, and to update it in real time.

It is an interesting research problem to define efficient synopsis data structures, that are cheap to create and maintain, but still contain sufficient details for query optimization.[12]

**PROJECTED ELUCIDATION:**

As the size of metadata might increase and will affect query execution cost, we can fragment metadata over server/ base station and fusion node (cluster head).

**Table 2.** Examples of WSN metadata elements for temperature data at base station

| Metadata Element | Value |
|---|---|
| Number of nodes | 21 |
| Network Id | NI3 |
| Location of WSN (Latitude & longitude) | Lat 40°26'North; Long 3°42'West |
| Phenomenon | Temperature |
| Data Unit | Celsius degree |

We store above metadata at base station which is required for query fragmentation.

Cluster head can have data to support query optimization as follows:

**Table 3.** Examples of WSN metadata elements for temperature data at cluster head

| Metadata Element | Value |
|---|---|
| NodeID | N3 |
| Mote type | Mica Mote |
| Sensor type | Sensirion SHT11 |
| Number of node neighborhood to the cluster | 7 |

Above two tables shows the *static* metadata maintained by the heterogeneous WSN but as data aggregated with timestamp from sensor over period form rows of the virtual sensor relation. As per user query size of attribute data-structure, memory location of data structure for virtual relation will describe the virtual metadata. It is updated as data is aggregated from the sensor nodes.

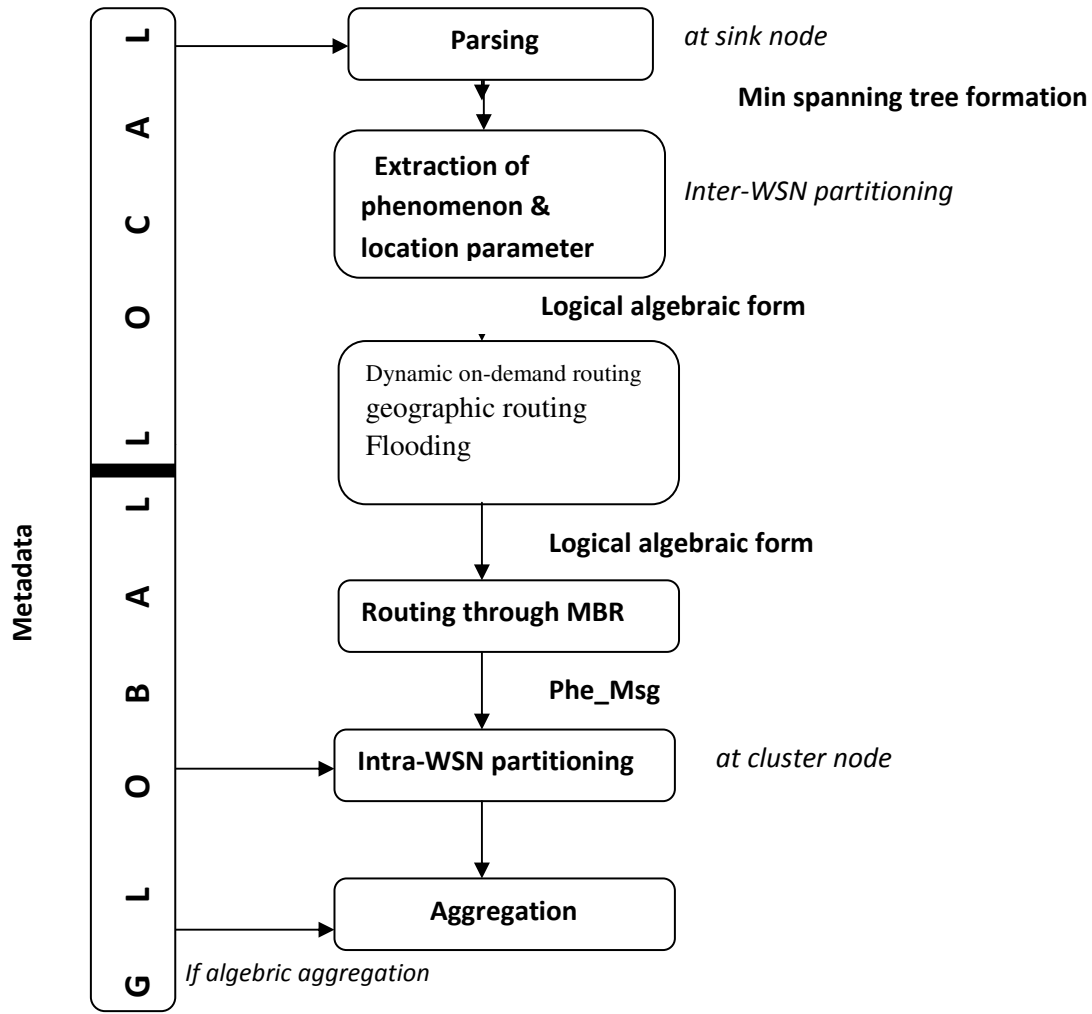

## 3.2 ISSUE: QUERY PLAN

Continuous queries can be categorized as follows:[11]

Query 1: "Return repeatedly the abnormal temperatures measured by all sensors."

Query 2: "Every minute, return the temperature measured by all sensors on the third floor".
Query 3: "Generate a notification whenever two sensors within 5 yards of each other simultaneously measure an abnormal temperature".
Query 4: "Every five minutes retrieve the maximum temperature measured over the last five minutes".
Query 5: "Return the average temperature measured on each floor over the last 10 minutes".

These example queries have the following characteristics:
- Monitoring queries are long running.
- The desired result of a query is typically a series of notifications of system activity (periodic or triggered by special situations).

- Queries need to correlate data produced simultaneously by different sensors.
- Queries need to aggregate sensor data over time windows.
- Most queries contain some condition restricting the set of sensors that are involved (usually geographical conditions).

Query plan for execution can follow one of the 3 strategies:

Queries are received at the base station and forwarded to the sensor network. Sensor network collect, filter, aggregate and send the response back to the server [1]. This is called *centralized or warehouse approach* to query processing.

In centralized models, query processing and access to sensor network were separated. The central server was the performance bottleneck and single point of failure. In addition, all sensors were required to send data to the central server, which incurred huge communication cost.

Enhanced centralized approaches move part of the processing to the sensor nodes. Enhanced centralized approach move part of the processing to the sensor node. By exploiting the limited computation capability of the sensors, these approaches successfully reduce the major cost on network communication. As a result, enhanced centralized or semi-distributed approaches such as Cougar and Fjord were proposed. Cougar and Fjord improve the centralized architecture by extracting only required data based on user queries.

In Cougar, sensor data is periodically collected from the physical environment and is represented by time series. Every measurement is associated with a timestamp. Other than sensor data, sensor attributes such as location are represented by relations and stored in the front-end server (static & global metadata) for querying. Fjord works on any language while Cougar provides a SQL-like language interface.

Now a days, smart sensors call mote sensors are capable not only of measuring real world phenomena but also filtering, sharing, and combining those measurements. Tiny Aggregation (TAG) and TinyDB are two distributed approaches that are significant to the development of query processing techniques on *ad-hoc* WSNs. Acquisitional query processing (ACQP) techniques are implemented in TinyDB which addresses the issues of when, where, and in what order the sensor nodes are sampled and which nodes should be included in processing a particular query. Distributed approach can be energy efficient when the query rate is less than the rate at which data is generated

TinyDB also includes support for grouped aggregation queries. Aggregation has the attractive property that it reduces the quantity of data that must be transmitted through the network; other sensor network research has noted that aggregation is perhaps the most common operation in the domain. Context aware hierarchical profiling of simple aggregation query to develop query plan.

**RESEARCH CHALLENGE:**

Hierarchical query partitioning to sub-queries, and each sub-query to be executed on selected WSN. More ever there may be heterogeneous classes of sensor nodes in a WSN to sense different attribute, therefore query is to be further partitioned for processing based on different classes of sensors in a WSN.

**PROJECTED ELUCIDATION**:

To overcome the problem of deciding the place for partitioning we believe that slightly different architecture is necessary to realize inter and intra query partitioning.

1. This architecture rests on two features. The first feature is *in-network* implementation of database operators through push computation. When a query is put to the sink node we parse query for location and phenomenon parameter to apply inter WSN query partitioning on the query.

2. Now query is disseminated as Phe_Msg across the network (either to all the nodes using simple flooding, or to geographically constrained set of nodes using variants of well-known geographic routing algorithms or reactive on-demand routing algorithm [15]) based on the global metadata stored at the sink.
    a. Phe_Msg is flooded through the network through MPRs. ID of the fusion node is passed along with the Phe_Msg. Phe_Msg message is discarded at the cluster head or if received in duplicate at any fusion node.
    b. When a node p sends a query request Phe_Msg to next node q in the network, it specifies the time slot in which it expects to hear from q. As soon as q receives Phe_Msg from p it synchronizes its time with p. Now q delivers the Phe_Msg to its descendent with delivery time such that q is ready with response before p expects to hear q.
    c. Each fusion node applies Bellman ford algorithm ( to support negative or reverse edges) in the reverse direction, from fusion node to sink. We can use Bellman algorithm because as a node/link break Err message is flooded through MBRs so count to infinity problem might not occur. Each node has full path to destination.

```
/ * Pseudo code of Modified Bellman Ford to get three shortest paths
from source to sink through reverse forwarding */
/* list vertices & edges are calculated from the path traveled by the ReqQ
msg */

proc Bellman Ford(list vertices, list edges, vertex source, vertex sink)

 /* Bellman ford step 1: Initialize graph */
for each vertex v in vertices:
    if v is sink then
            v.distance := 0
    else
            v.distance := infinity
    v.predecessor := null
```

```
        /* Bellman ford step 2: relax edges repeatedly */
        For p=1 to 3

        /* p refers to p^th shortest path
          For i = 1 to size(vertices)-1:
                    For each edge uv in edges :

                    /* uv is the edge from u to v */
                    If node u.flag=1 and v.flag=1 then

                       /* Vertices has flag 1 if they have
                       received & retained Phe_Msg msg */
        u := uv.source
                   v := uv.destination
                      If u.distance + uv.weight <
                        v.distance:
                                v.distance := u.distance + uv.weight
                                v.predecessor := u
                    next                                            edge
          next i
        /*Bellman ford step 3: check for negative-weight cycles */
        For each edge uv in edges:
          u := uv.source
          v := uv.destination
          If u.distance + uv.weight < v.distance:
                error "Graph contains a negative-weight cycle"

          loc = 0
          v=source
          While v != sink
             Route[p][loc++]= v

             /* Route is 2-D array where 1st dimension is route
             number & 2nd dimension is route trace */

             delete_edge(v, v.predecessor )

             /* delete_edge eliminates edge v,v.predecessor
             from the list edges */
        v = v.predecessor

Next p
```

    d. If one path from sink to cluster head is traversed by Phe_Msg then all fusion nodes on the path will have 3 best route to the cluster head, which will help them next time onwards when Phe_Msg is flooded after inter-WSN query partitioning. Phe_Msg will be discarded at the earliest fusion node if that fusion node does not lead to the WSN on which the sub-query / Phe_Msg resulted from inter-WSN partitioning is to be processed.
3. When Phe_Msg reached at cluster head, it applies intra-WSN partitioning (for heterogeneous WSN where one WSN cluster may have sensor with different phenomenon) so that based on the value of phenomenon parameter in Phe_Msg it is directed to particular sensor based on the local metadata stored at the cluster node.
4. In response to the query, each node generates tuples that match the query, and transmits matching tuples towards the origin of the query. As the tuples are routed through the network, intermediate nodes might apply one or more database operators.
5. A query may contain a grouping clause. In this case, the root inserts the grouping condition into the query request to be routed down the network. Query execution follows the approach described above except that partial state records are tagged with a group ID. Any leaf node applies the grouping expression to compute a group ID, tags its record with this group and sends it to its parent. When a node receives a record from a child, it compares its own group ID with the one that is in its child's record. If they coincide, the node combines the two values according to the aggregation function specified in the query. Otherwise, the node stores its own value and the received one separately. If another child's message arrives with a value in either group, the node updates the right aggregate.
6. Cluster head calculate aggregate over the reading from sensors of the cluster and as data is moved up the minimum spanning tree in response to Phe_Msg aggregate values are routed up to sink.

For the example query stated in section 2.2.1 below is the hierarchical, inter and intra partitioning.

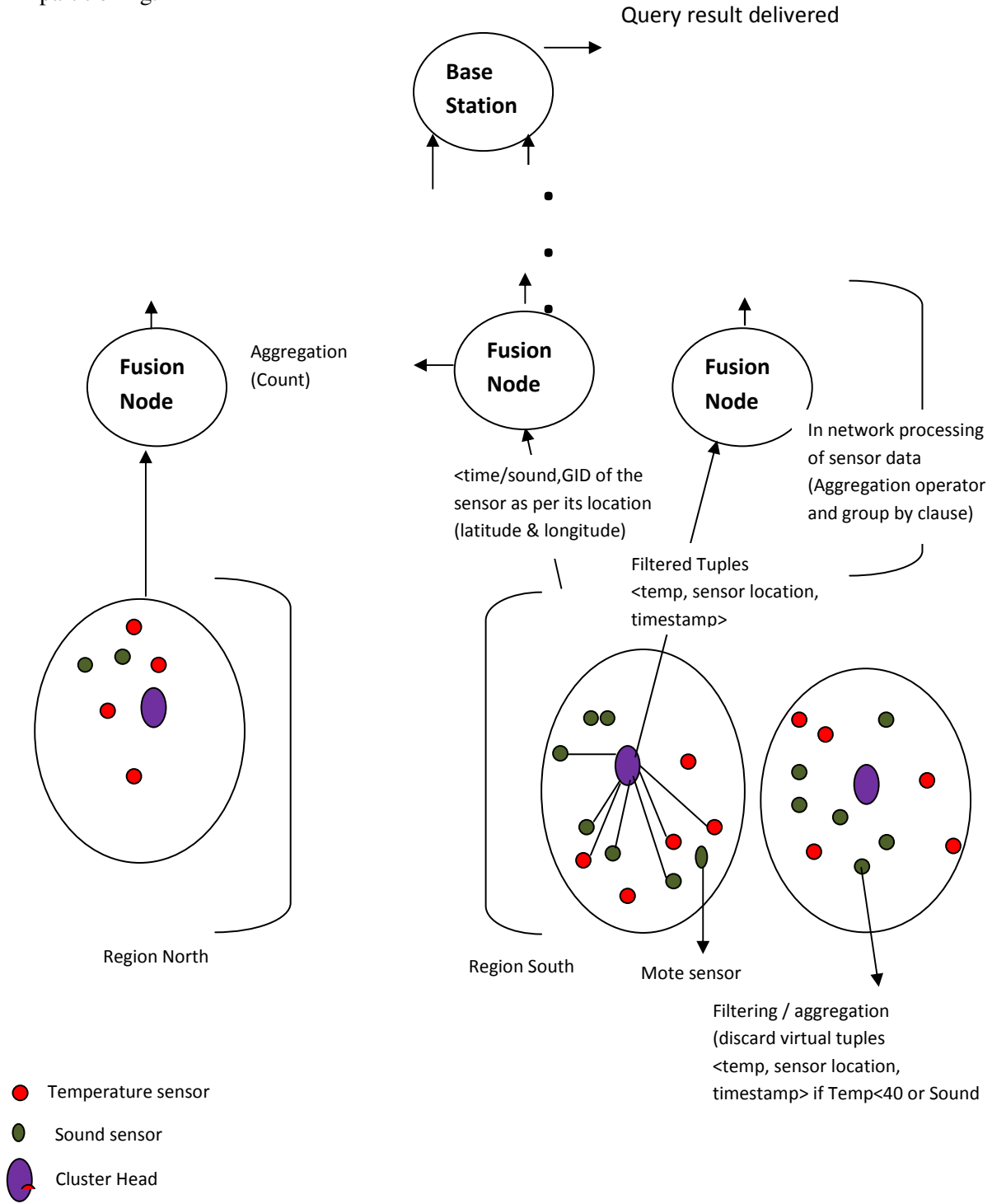

## 4. CONCLUSION

Energy constraint in wireless sensor network makes the challenges in query processing an important research area. We focus on application layer issues which contribute to the optimization of the query processing. We studied the performance of query processing engines (TinyDB, Cougar) from the perspective of location of metadata and query processing semantics. Centralized model of query processing can make accurate query plan but communication cost for metadata collection is bottleneck. Inter-WSN and intra-WSN partitioning of query with in-network processing of query processing operators along the minimum spanning tree routed at the sink node can reduce communication cost involved in metadata collection. We consider the case where queries are synchronously injected at the sink node and query is single long running aggregate query. In this paper, we address the problem of centralized metadata and formulate a solution based on distributed metadata. In the future we plan to extend our approach to support multiple query optimizations. Additionally, we plan to calculate of cost at different levels of wireless sensor network. This cost includes metadata collection, sampling cost, reporting cost and building structure cost.